# Optical Wireless Satellite Networks versus Optical Fiber Terrestrial Networks: The Latency Perspective

*Invited Paper*

Aizaz U. Chaudhry and Halim Yanikomeroglu
Department of Systems and Computer Engineering, Carleton University, Ottawa, Canada – K1S 5B6
{auhchaud, halim}@sce.carleton.ca

*Abstract*—Formed by using laser inter-satellite links (LISLs) among satellites in upcoming low Earth orbit and very low Earth orbit satellite constellations, optical wireless satellite networks (OWSNs), also known as free-space optical satellite networks, can provide a better alternative to existing optical fiber terrestrial networks (OFTNs) for long-distance inter-continental data communications. The LISLs operate at the speed of light in vacuum in space, which gives OWSNs a crucial advantage over OFTNs in terms of latency. In this paper, we employ the satellite constellation for Phase I of Starlink and LISLs between satellites to simulate an OWSN. Then, we compare the network latency of this OWSN and the OFTN under three different scenarios for long-distance inter-continental data communications. The results show that the OWSN performs better than the OFTN in all scenarios. It is observed that the longer the length of the inter-continental connection between the source and the destination, the better the latency improvement offered by the OWSN compared to OFTN.

*Keywords—free-space optical satellite networks, laser inter-satellite links, network latency, optical fiber terrestrial networks, optical wireless satellite networks, Starlink.*

## I. Introduction

Laser inter-satellite links (LISLs) [1] between satellites in upcoming low Earth orbit (LEO) or very low Earth orbit (VLEO) satellite constellations will enable the creation of optical wireless satellite networks (OWSNs) among satellites in these constellations. These satellite networks are also referred to as free-space optical satellite networks. The LISLs will be crucial in ensuring low-latency paths within the OWSN. In their absence, a long-distance inter-continental connection over the OWSN between two cities, such as New York and Dublin, will have to go back and forth between grounds stations and satellites, which will negatively impact network latency.

The primary use case of an OWSN that is formed by LISLs in upcoming LEO/VLEO satellite constellations, like Phase I of SpaceX's Starlink [2], could be the provision of long-distance low-latency communications. An OWSN can provide low-latency communications as a premium service to the financial hubs around the world, and this use case can easily recover the cost of deploying and sustaining such an OWSN. A one millisecond advantage can translate into $100 million a year in revenues for a major brokerage firm in trading stocks at the stock exchange [3], and an advantage of a few milliseconds in financial stock markets may mean billions of dollars of revenues for these financial firms. Technological solutions, such as low-latency communications networks, are being highly sought after by these firms, and a low-latency OWSN could be the perfect solution.

Unlike optical communications in optical fiber terrestrial networks (OFTNs) where data is sent using an optical carrier (i.e., a laser beam) over a guided medium like optical fiber, data in OWSNs is sent over LISLs by employing laser beams between satellites over the vacuum in space – an unguided medium. The ratio of the speed of light in vacuum to the speed of light in a medium is called the refractive index or index of refraction of that medium [4]. A higher refractive index for a medium translates into slower transmission of light through that medium. The light will travel through a medium at 1/2 the speed of light in vacuum if it has a refractive index of 2. The speed of light in vacuum is usually denoted by $c$ [5].

Optical fibers are generally made of glass and have an index of refraction of approximately 1.5. This indicates that the speed of light in optical fibers is approximately $c/1.5$, which means that the speed of light in vacuum is approximately 50% higher than the speed of light in optical fiber. This has critical significance in OWSNs. The higher speed of light in vacuum gives OWSNs a critical advantage over OFTNs in terms of latency for long-distance data communications.

The end-to-end delay in the network from source to destination comprises transmission delay, processing delay, queueing delay, and propagation delay [6]. In OFTNs or OWSNs, propagation delay means the delay caused by the transmission of the optical signal along the medium, i.e., optical fiber or vacuum. The propagation delay is directly related to the end-to-end distance between the source and the destination, and gets very significant for long-distance data communications [7]. In this paper, we investigate the network latency of OWSNs versus OFTNs, and we define latency (or network latency) as the propagation delay from the source to the destination. For this comparison, we employ Starlink's Phase I constellation, and assume LISLs between satellites in this constellation to realize an OWSN. We consider three different scenarios for long-haul inter-continental data communications, including connections between New York and Dublin, Sao Paulo and London, and Toronto and Sydney. We find minimum-latency paths between cities over the two networks for each scenario. The results indicate that the OWSN outperforms the OFTN in terms of latency. The greater the inter-continental distance between cities, the higher the improvement in latency with OWSN compared to OFTN.

The rest of the paper is organized as follows. Section II briefly discusses the related work. The details of different steps of our methodology for calculating latency of an inter-continental connection between cities over the OFTN and the OWSN are given in Section III. Section IV presents the results of the comparison of these two networks in terms of latency. Conclusions and future work are provided in Section V.

## II. RELATED WORK

Creating an OWSN by using LISLs within a satellite constellation and the problem of routing over this network has been examined [8]. It is stated that an OWSN can provide lower latency communications than an OFTN for data communications over long distances of more than 3,000 km.

Using a hypothetical constellation of 1,600 LEO satellites at 550 km altitude at an inclination of 53º with respect to the Equator in 40 orbital planes (OPs) with 40 satellites in each OP, a median round trip time improvement of 70% with the satellite network was found when comparing with Internet latency [9]. However, this comparison was overly favorable to the satellite network as delays due to sub-optimal routing, congestion, queueing, and forward error correction were not accounted for in the satellite network, while such delays were considered in measuring Internet latency. In our work, we compare OWSN and OFTN in terms of propagation delay only. In fact, our comparison may be favorable to the OFTN as we consider the shortest distance between two cities over the OFTN along the surface of the Earth. This is not the case in reality, and the OFTN may not provide the shortest path along Earth's surface between cities in different continents. For example, instead of following the shortest path to connect two points on Earth's surface, long-haul submarine optical fiber cables are laid along paths that avoid earthquake prone areas and difficult seabed terrains with high slopes [10]. Also, instead of a hypothetical constellation, we employ the satellite constellation for Phase I of Starlink for this comparison.

In an earlier work [11], we investigated a use case for OWSNs to check their suitability for providing low-latency communications over long distances as their primary service. It was shown that an OWSN operating at 550 km altitude outperforms an OFTN in terms of latency when communication distances are greater than 3,000 km. In this paper, we build on our work in [11] and investigate the network latency of OWSNs versus OFTNs under more realistic scenarios for long-distance inter-continental data communications.

## III. METHODOLOGY

### A. Optical Fiber Terrestrial Network

The refractive index of a single-mode optical fiber (manufactured by Corning®) that is suitable for long-distance communications is 1.4675 at 1,310 nm operating wavelength [12]. Consequently, the speed of light in this optical fiber is $c/1.4675$. The exact value of the speed of light in vacuum or $c$ is 299,792,458 m/s [13]. We consider the speed of light in optical fiber in the OFTN to be $c/1.4675$, i.e., 204,287,876 m/s.

The latency of an inter-continental long-distance connection between two cities over the OFTN along the surface of the Earth is calculated by dividing the shortest distance between the two cities with this speed of light in optical fiber. The shortest distance between two cities is calculated by using their coordinates (latitudes and longitudes) on the surface of the Earth. The radius of the Earth is considered as 6,378 km for this calculation. For example, the shortest distance between New York–Dublin, Sao Paulo–London, and Toronto–Sydney inter-continental connections along the surface of the Earth is calculated as 5,121.30 km, 9,514.30 km, and 15,584.58 km, respectively. For this purpose, we use the coordinates of the financial stock markets (i.e., New York Stock Exchange, Dublin Stock Exchange, Sao Paulo Stock Exchange, London Stock Exchange, Toronto Stock Exchange, and Sydney Stock Exchange) within these cities. Fig. 1 illustrates the shortest distance path in yellow color between New York and Dublin over the OFTN along the Earth's surface.

### B. Optical Wireless Satellite Network

To investigate the network latency of OWSNs versus OFTNs, we employ the satellite constellation for Phase I of Starlink. This constellation will have a total of 1,584 LEO satellites that will be deployed at 550 km altitude at an inclination of 53º with respect to the Equator in 24 OPs and each OP will have 66 satellites [2]. When assuming this constellation to be uniform, the spacing between OPs is 15º and the spacing between satellites within an OP is 5.45º.

LISLs are assumed between satellites in this constellation to realize an OWSN. The LISL range for all satellites is assumed to be 1,500 km, and the LISL range is defined as the distance over which a satellite can form a laser inter-satellite link with another satellite. The resulting connectivity for a satellite (i.e., satellite *x10101* in this example) in the OWSN is shown in Fig. 2. It consists of four intra-OP neighbors (i.e., four neighbors in the same OP within the LISL range including two in the front, i.e., *x10102* and *x10103*, and two at the rear, i.e., *x10165* and *x10166*), two adjacent OP neighbors (i.e., two nearest neighbors within the LISL range in right and left OPs, i.e., *x10265* and *x12454*), and six crossing OP neighbors within the LISL range in three different crossing OPs (i.e., *x11229* in crossing OP *x112*; *x11325*, *x11326*, *x11327*, and *x11328* in crossing OP *x113*; and *x11424* in crossing OP *x114*).

We consider the exact value of the speed of light in vacuum for calculating latency in the OWSN. The latency of an LISL in the OWSN is calculated by dividing the length of that link (i.e., the distance between satellites at the two ends of that link) by the speed of light in vacuum. Then, Dijkstra's shortest path algorithm [14] is employed to find the shortest path between two cities over the OWSN in terms of link latency, which is in fact the minimum-latency route between cities over the OWSN. It includes the latency of the optical link from the ground station (GS) in the source city on Earth to the ingress satellite of the OWSN in space, the latencies of the LISLs in this path, and the latency of the optical link from the egress satellite of the OWSN in space to the GS in the destination city on Earth. The time is divided into time slots of 1 second duration. A shortest path (i.e., a route with minimum latency) is calculated at each time slot where a time slot represents a snapshot of the OWSN at that second.

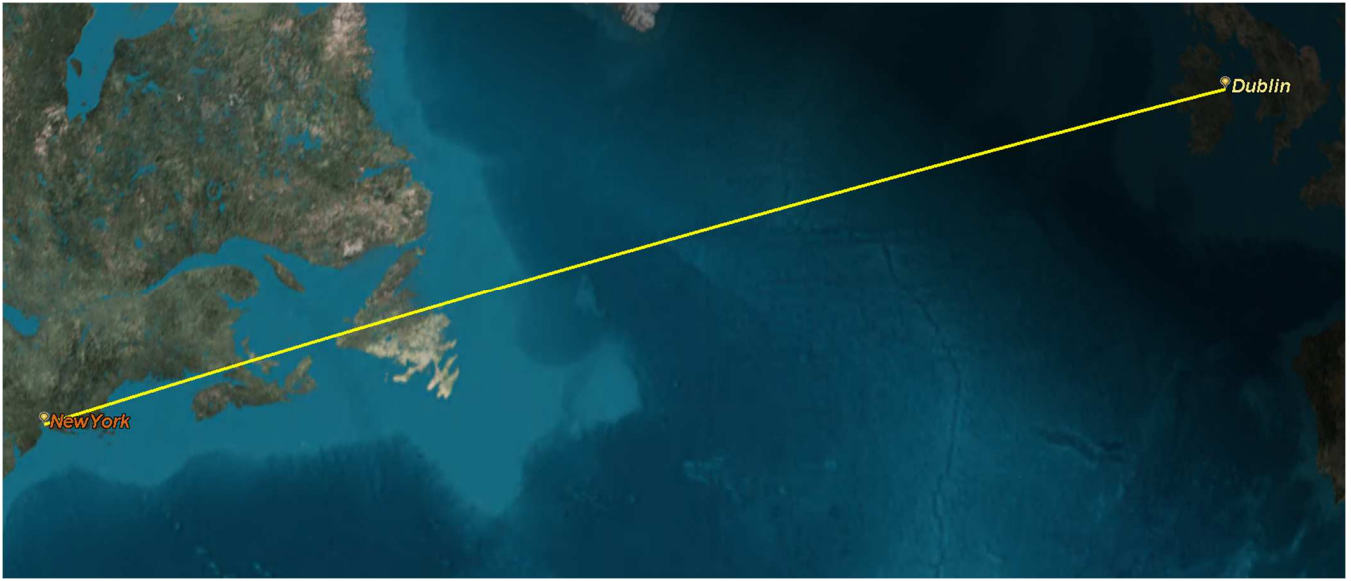

Fig. 1. Shortest distance path between New York and Dublin over the OFTN.

Compared to OFTN, an extra distance from a GS on Earth to ingress satellite and from egress satellite to a GS is involved when communicating over the OWSN. However, the extra latency due to this extra distance becomes insignificant for long-distance inter-continental communications over the OWSN.

IV. RESULTS

We compare OWSN and OFTN in terms of latency in three different inter-continental connection scenarios, including New York–Dublin, Sao Paulo–London, and Toronto–Sydney. To simulate the OWSN, we use the well-known satellite constellation simulator STK Version 12.1 [15]. We create the satellite constellation for Phase I of Starlink in STK using the parameters discussed in Section III.

Distinct IDs are generated for the 1,584 satellites within this constellation. The following IDs are created for the 24 OPs: {$x101$, $x102$, $x103$, …, $x124$}. For the 66 satellites within each OP, distinct IDs are created, e.g., the following IDs are created for satellites in the first OP: {$x10101$, $x10102$, $x10103$, …, $x10166$}.

The satellites in this constellation travel at speeds of approximately 7.6 km/s. This means that the GS to ingress satellite link, satellite to satellite links, and egress satellite to GS link or the latencies of these links are constantly changing. Consequently, the shortest path of the inter-continental connection over the OWSN between two cities (and/or its latency) also changes at every time slot.

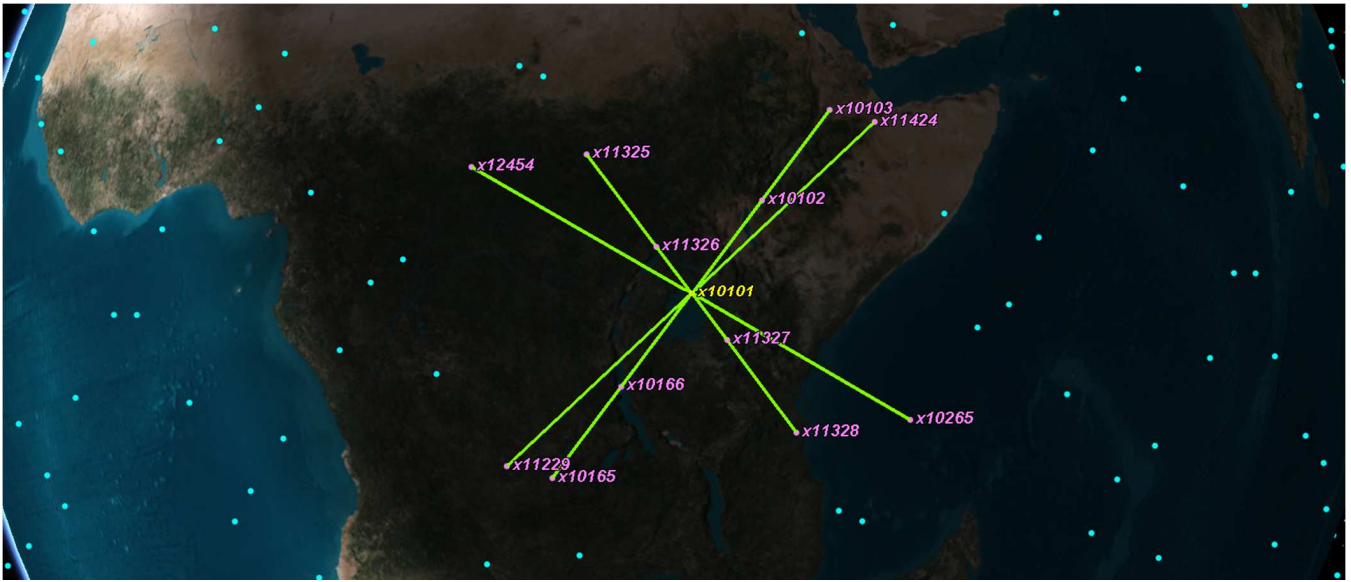

Fig. 2. Satellite connectivity in the OWSN.

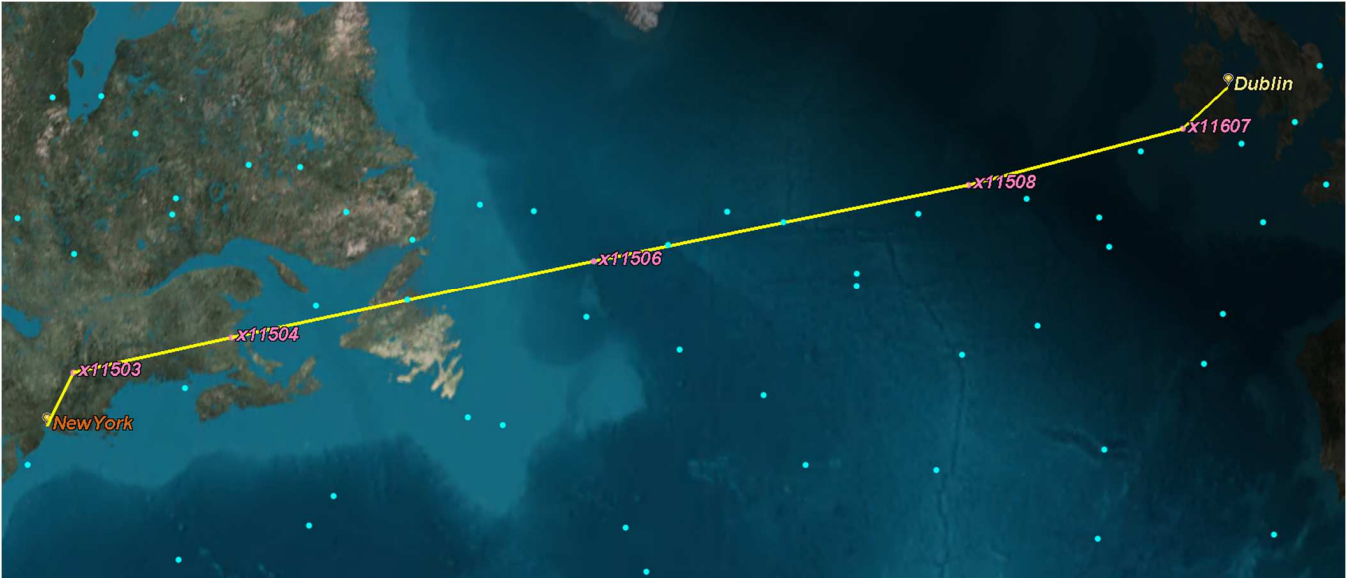

Fig. 3. Shortest path between New York and Dublin over the OWSN at 1st time slot.

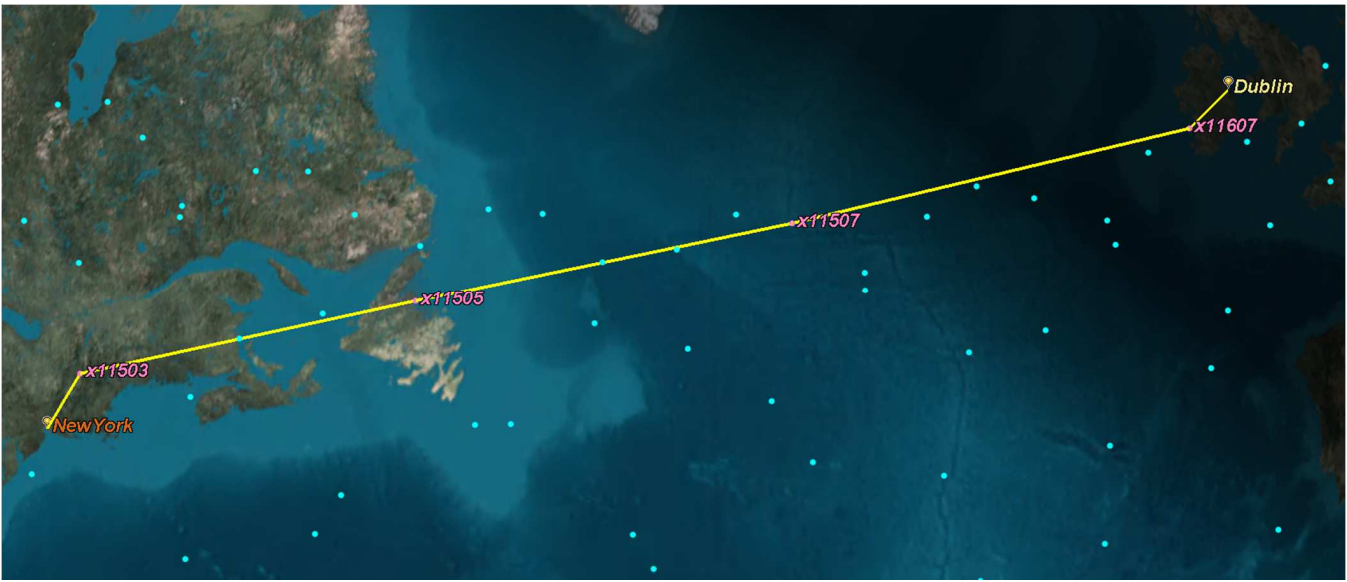

Fig. 4. Shortest path between New York and Dublin over the OWSN at 5th time slot.

We run the simulation for one hour and find the shortest path in terms of latency between two cities over this OWSN at every second (or time slot). For example, the shortest paths calculated at the first twenty time slots for the New York–Dublin inter-continental connection and their latencies are given in Table 1. Furthermore, the shortest path for this inter-continental connection over the OWSN at first time slot is shown in Fig. 3 in yellow. This shortest path consists of the GS at New York (located at the New York Stock Exchange), satellites *x11503* (ingress), *x11504*, *x11506*, *x11508*, and *x11607* (egress) in the OWSN, and the GS at Dublin (located at the Dublin Stock Exchange). The latency of this shortest path decreases over the next three time slots as indicated in Table 1. Note that a shortest path with lower latency becomes available at the fifth time slot, and this shortest path is shown in Fig. 4. The latency of this path increases at succeeding time slots, and at eleventh time slot, a shortest path with lower latency is found as can be seen from Table 1. This shortest path is shown in Fig. 5; it continues to exist till eighteenth time slot after which it becomes unavailable and a new shortest path with a higher latency is established at the nineteenth time slot. Due to the high-speed movement of satellites along their OPs, the shortest path and/or its latency change at every time slot.

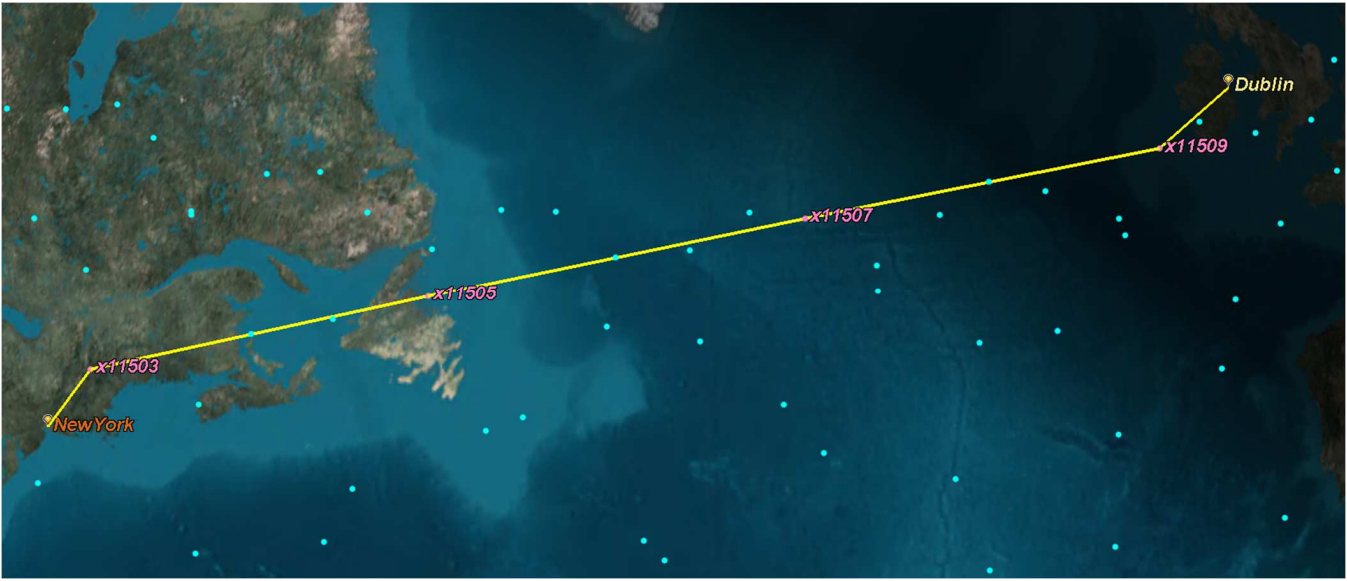

Fig. 5. Shortest path between New York and Dublin over the OWSN at 11th time slot.

Table 2 shows the results of the comparison of OWSN and OFTN in terms of latency in milliseconds. The latency of an inter-continental connection over the OWSN shown in this table is the average of the latencies of all shortest paths that are calculated at all time slots over the simulation duration of one hour.

Compared to OFTN, the latency improvement with the OWSN is 5.00 ms, 9.93 ms, and 17.95 ms for New York–Dublin, Sao Paulo–London, and Toronto–Sydney long-distance connections, respectively. These results clearly indicate that the OWSN outperforms the OFTN in all scenarios in terms of network latency. In fact, the longer the length of the inter-continental connection between a pair of cities, the higher the improvement in latency provided by the OWSN compared to OFTN.

## V. CONCLUSIONS

We examine the network latency of OWSN versus OFTN under three different scenarios for long-distance inter-continental data communications. We simulate the OWSN by using LISLs among satellites in the constellation for Phase I of Starlink. We find the shortest path between two cities in terms of link latency (i.e., the minimum-latency route between cities) over the OWSN at every time slot. Then, we compare the average latency of shortest paths at all time slots in the OWSN with latency of the long-distance inter-continental connection between these cities over the OFTN along the Earth's surface. In all scenarios, the OWSN performs better than the OFTN in terms of latency, and provides an improvement of 5.00 ms, 9.93 ms, and 17.95 ms (or 19.94%, 21.32%, and 23.53%) for New York–Dublin, Sao Paulo–London, and Toronto–Sydney long-distance inter-continental connections, respectively.

We discover an interesting relationship between network latency over the OWSN and length of the inter-continental connection between cities. We find that the longer the inter-continental connection, the greater the gain due to the higher speed of light over free-space laser links in OWSN compared to the speed of light over optical fiber laser links in OFTN, and the better the improvement in network latency offered by the OWSN compared to OFTN. Based on these findings, we can say that OWSNs resulting from LISLs in upcoming LEO/VLEO satellite constellations, like Phase I of Starlink, can be the perfect solution for high-speed trading firms seeking low-latency inter-continental data communications among financial stock markets around the world.

In this work, we employed a LISL range of 1,500 km for all satellites while calculating latency of an inter-continental connection for data communications over the OWSN. Different LISL ranges may impact latency differently, and as part of future work, we intend to investigate the effect of different LISL ranges on the latency of the OWSN.

## ACKNOWLEDGEMENT

This work has been supported by the National Research Council Canada's (NRC) High Throughput Secure Networks program (CSTIP Grant #CH-HTSN-625) within the Optical Satellite Communications Consortium Canada (OSC) framework. The authors would like to thank AGI for the STK platform.

## REFERENCES

[1] A.U. Chaudhry and H. Yanikomeroglu, "Laser Intersatellite Links in a Starlink Constellation: A Classification and Analysis," *IEEE Vehicular Technology Magazine*, vol. 16(2), pp. 48–56, Jun. 2021.

[2] SpaceX FCC update, 2018, "SpaceX Non-Geostationary Satellite System, Attachment A, Technical Information to Supplement Schedule S," [Online]. Available: https://licensing.fcc.gov/myibfs/download.do?attachment_key=1569860.

[3] R. Martin, "Wall Street's Quest to Process Data at the Speed of Light," *Information Week*, Apr. 2007, [Online]. Available: https://www.informationweek.com/wall-streets-quest-to-process-data-at-the-speed-of-light/d/d-id/1054287.

[4] S.C. Gupta, *Textbook on Optical Fiber Communication and Its Applications*, PHI Learning Private Limited, New Delhi, 2018.

[5] J. Hecht, *Understanding Lasers: An Entry-Level Guide*, Wiley, Hoboken, NJ, 2018.

[6] J.F. Kurose and K.W. Ross, *Computer Networking – A Top Down Approach Featuring the Internet*, Addison-Wesley, Boston, MA, 2009.

[7] Corning Inc., "The Super Connected World: Optical Fiber Advances and Next Gen Backbone, Mobile Backhaul, and Access Networks [White Paper]," Jun. 2012, [Online]. Available: https://www.corning.com/media/worldwide/coc/documents/Fiber/white-paper/wp6024.pdf.

[8] M. Handley, "Delay is Not an Option: Low Latency Routing in Space," in *Proc. 17th ACM Workshop on Hot Topics in Networks*, Redmond, WA, USA, 2018, pp. 85–91.

[9] D. Bhattacherjee and A. Singla, "Network Topology Design at 27,000 km/hour," in *Proc. 15th International Conference on Emerging Networking Experiments And Technologies*, Orlando, FL, USA, 2019, pp. 341–354.

[10] Z. Wang, Q. Wang, B. Moran, and M. Zukerman, "Terrain Constrained Path Planning for Long-Haul Cables," *Optic Express*, vol. 27(6), pp. 8221–8235, Mar. 2019.

[11] A.U. Chaudhry and H. Yanikomeroglu, "Free Space Optics for Next-Generation Satellite Networks," *IEEE Consumer Electronics Magazine (Early Access)*, doi: 10.1109/MCE.2020.3029772.

[12] Corning Inc., "Corning® Hermetic Single-mode and Multimode Specialty Optical Fibers," Mar. 2010, [Online]. Available: https://www.corning.com/media/worldwide/csm/documents/M0300013v00_Hermetic_Acrylate_March_2010.pdf.

[13] R.J. MacKay and R.W. Oldford, "Scientific Method, Statistical Method and the Speed of Light," *Statistical Science*, vol. 15(3), pp. 254–278, Aug. 2000.

[14] E.W. Dijkstra, "A Note on Two Problems in Connexion with Graphs," *Numerische Mathematik*, vol. 1, pp. 269–271, Dec. 1959.

[15] AGI, "Systems Tool Kit (STK)," [Online]. Available: https://www.agi.com/products/stk.

TABLE 1. SHORTEST PATHS AT FIRST TWENTY TIME SLOTS OVER THE OWSN FOR NEW YORK–DUBLIN INTER-CONTINENTAL CONNECTION

| Time Slot (s) | Shortest Path | Latency (ms) |
|---|---|---|
| 1 | GS at New York, satellites *x11503, x11504, x11506, x11508, x11607*, GS at Dublin | 19.8152 |
| 2 | GS at New York, satellites *x11503, x11504, x11506, x11508, x11607*, GS at Dublin | 19.8148 |
| 3 | GS at New York, satellites *x11503, x11504, x11506, x11508, x11607*, GS at Dublin | 19.8145 |
| 4 | GS at New York, satellites *x11503, x11504, x11506, x11508, x11607*, GS at Dublin | 19.8144 |
| 5 | GS at New York, satellites *x11503, x11505, x11507, x11607*, GS at Dublin | 19.8042 |
| 6 | GS at New York, satellites *x11503, x11505, x11507, x11607*, GS at Dublin | 19.8042 |
| 7 | GS at New York, satellites *x11503, x11505, x11507, x11607*, GS at Dublin | 19.8044 |
| 8 | GS at New York, satellites *x11503, x11505, x11507, x11607*, GS at Dublin | 19.8047 |
| 9 | GS at New York, satellites *x11503, x11505, x11507, x11607*, GS at Dublin | 19.8052 |
| 10 | GS at New York, satellites *x11503, x11505, x11507, x11607*, GS at Dublin | 19.8058 |
| 11 | GS at New York, satellites *x11503, x11505, x11507, x11509*, GS at Dublin | 19.6848 |
| 12 | GS at New York, satellites *x11503, x11505, x11507, x11509*, GS at Dublin | 19.6851 |
| 13 | GS at New York, satellites *x11503, x11505, x11507, x11509*, GS at Dublin | 19.6854 |
| 14 | GS at New York, satellites *x11503, x11505, x11507, x11509*, GS at Dublin | 19.6859 |
| 15 | GS at New York, satellites *x11503, x11505, x11507, x11509*, GS at Dublin | 19.6865 |
| 16 | GS at New York, satellites *x11503, x11505, x11507, x11509*, GS at Dublin | 19.6872 |
| 17 | GS at New York, satellites *x11503, x11505, x11507, x11509*, GS at Dublin | 19.6880 |
| 18 | GS at New York, satellites *x11503, x11505, x11507, x11509*, GS at Dublin | 19.6890 |
| 19 | GS at New York, satellites *x10918, x11114, x11311, x11508, x11509*, GS at Dublin | 20.2957 |
| 20 | GS at New York, satellites *x10918, x11114, x11311, x11508, x11509*, GS at Dublin | 20.3069 |

TABLE 2. LATENCY – OWSN VS. OFTN

| Inter-Continental Connection | Latency (ms) | | Latency Improvement | |
|---|---|---|---|---|
| | OFTN | OWSN | (ms) | (%) |
| **New York–Dublin** | 25.07 | 20.07 | 5.00 | 19.94 |
| **Sao Paulo–London** | 46.57 | 36.64 | 9.93 | 21.32 |
| **Toronto–Sydney** | 76.29 | 58.34 | 17.95 | 23.53 |